# WiFiPos: An In/Out-Door Positioning Tool

## Juan Toloza[1], Nelson Acosta, Carlos Kornuta[2]

[1] *(Post-Doctoral Fellow, CONICET, INCA/INTIA - School of Exact Sciences – UNICEN, TANDIL – Argentina)*
[2] *(Post-Doctoral Fellow, CONICET, INCA/INTIA - School of Exact Sciences – UNICEN, TANDIL – Argentina)*

***ABSTRACT:*** *Geographical location solutions have a wide diversity of applications, ranging from emergency services to access to tourist and entertainment services. GPS (Global Positioning System) is the most widely used system for outdoor areas. Since it requires a direct line of sight with the satellites, it cannot be used indoors. For indoor positioning, the most commonly used technology to calculate the position of mobile devices is Wi-Fi. In this article, we present a positioning tool called WiFiPos based on Wi-Fi signal processing. This tool uses different variations of the Fingerprint technique to analyze the performance and accuracy of the system both indoors and outdoors.*

## I. Introduction

Current Location Based Services (LBS) can be classified as satellite-based (GNSS, Global Navigation Satellite System) and Wireless Network-based (cellular networks, Wi-Fi or Ultra Wide Band), or hybrid systems that use techniques from both technologies. LBS provide users with applications for vehicle navigation, fleet management, emergency service identification, environment monitoring, and so forth.

The GPS system offers an accuracy of 10-15 meters [1], although there are various criteria that can be applied to improve it [2]. It has a major disadvantage, and that is that the error increases significantly in urban spaces and it is practically non-operational and cannot be used in enclosed environments, because it requires a clear, unobstructed line of sight between the device and a minimum of three satellites [2] [3].

Access Points (APs) to a Wi-Fi network can be used as reference points to calculate the position of a mobile device (MD). In this sense, positioning systems that are based on Wi-Fi signals have gained great significance in recent years because they allow calculating the position of a MD using an already-existing infrastructure in most buildings and public places, meaning that the installation of no additional hardware is required on the MDs, since most of them already offer access to Wi-Fi connections. These systems offer an accuracy of 2-3 meters in enclosed spaces, and 10 meters in outdoor areas [4] [5].

The parameter that is most commonly used as sensing technique is signal strength (RSSI, Received Signal Strength Indicator). Using this technique, there are two main models to calculate the position of a MD: signal propagation model and empirical model [6] [7] [8]. The former is based on the application of mathematical models that determine signal behavior and its degradation while it propagates. The empirical model calculates the position based on the relation between the actual values and the parameters that are stored in the database. Within this model, the algorithm that is most widely used for calculating positions is the Fingerprint algorithm.

*WiFiPos* is a positioning tool that uses the information from LAN 802.11 networks [9]. The system processes the RSSI parameter of the frames transmitted by the APs to calculate the location of the MD. *WiFiPos* calculates the position of the device by implementing variations to the Fingerprint technique, using various statistical and mathematical algorithms to analyze system accuracy.

This paper is organized as follows: Section 2 describes the state of the art, Section 3 details the design and the various components of the tool, Section 4 presents the experiments carried out, Section 5 discusses the results after implementing the different techniques, and Section 6 assesses the performance of the tool. Finally, Section 7 describes the conclusions and future work.

## II.    State of the Art

This section has been divided in two parts or subsections – the first one will present some Wi-Fi positioning systems for enclosed spaces, whereas the second part will focus on the analysis of the features offered by some of the Wi-Fi positioning tools that are based on the Fingerprint technique, implemented on standalone architectures.

Within the empirical model, most positioning systems use triangulation or fingerprinting algorithms to calculate the position of the MD. In the case of the Fingerprint algorithm, first a radio map is designed containing RSSI measurements for each visible AP in the coverage area of the MD. After all values have been obtained, the mode or average is calculated with the purpose of grouping the values, obtaining a strengths vector for each sampling point. These vectors are stored in the Fingerprint database. Thus, the location of any device can be determined by comparing the strengths vector obtained against the vectors stored [10].

Several developments have been carried out in the area of device location using the Fingerprint technique. One of the first approaches in this context is the RADAR system [6], which combines two methods – an empirical model using Fingerprint and a mathematical model that takes into account signal propagation. It obtains a mean accuracy within 2-3 meters. In [7], the authors of the RADAR system implement an improvement aimed at analyzing and reducing signal-inherent problems, such as multipath and interference, and analyze the environmental changes produced during the experimental phase. The accuracy obtained is less than 2 meters. In 2002, the Fingerprint technique is used to calculate the position of a MD in combination with a probabilistic model [11], which uses a technique that calculates the probability that the device is at a certain position within the radio map, based on Bayes algorithm. The system achieves an accuracy of 1.5-3 meters.

In 2010, the Fingerprint technique is used in combination with a propagation model [12] to calculate the position of a MD. The propagation model is defined based on the physical features of the environment, calculating the Wall Attenuation Factor (WAF). The proposed system uses a filter to improve accuracy, and achieves an error that is below 1.8 meters.

As already mentioned, the GPS system is not efficient in urbanized spaces because there are "shadowed areas" that are not reached by the signal from the satellites. In this context, one of the alternatives to be used are systems based on Wi-Fi signals.

In 2005, the University of California, San Diego (UCSD) carried out one of the first research works in this area with the purpose of locating a MD inside and outside campus facilities. The project was called Active Campus Project. It uses a context-based positioning technology that presents an interactive map of the place to the user of the MD and, through interaction with it, builds the positioning database. Then, the project was redesigned with the name of the Place Lab software in order to calculate positions at a metropolitan scale. It is based on the use of three technologies – Wi-Fi, GSM and Bluetooth – to calculate the position of the MD; the system achieves an accuracy of 15-20 meters. In the context of the Place Lab project, the authors in [13] analyze the performance of the Place Lab system and a suite of several algorithms (Fingerprint, Bayesian Filters, Centroid) implemented in a Wi-Fi positioning system and applied at a metropolitan scale. During the experimental phase, the techniques are tested in 3 different scenarios; the Fingerprint algorithm achieves a mean error of 13-20 meters.

In 2006, the MARA University of Technology in Malaysia [14] developed an outdoor positioning system based on the Fingerprint technique using the KNN (K nearest neighbors) algorithm as metric; the system achieves an accuracy of 12 meters.

There are several commercial systems that use the Fingerprint technique. LEASE [15] achieves an accuracy of 2.1 meters. The EKAHAU commercial system [16] is a positioning software that consists in an administrator, a server, and a client. System accuracy is 2-3 meters. The HORUS system, proposed by the authors of [17], achieves a high accuracy; it is based on measurements of the RSSI parameters, considering various environmental factors, and it calculates positions based on probabilistic algorithms and clustering techniques. The system has a relative error of 0.5 meters.

## III.    The Tool

The tool allows analyzing data from several APs by applying various techniques for accurately positioning a MD. The tool has two modes: online, which allows instantly positioning the MD, and offline, which allows processing the data collected by applying a set of techniques. The tool is designed as presented in [18], so that in the future they can complement each other.

The first process consists in creating a database, known as radio map, where all data collected are stored. In this sense, two structures are created for data storage – one that allows storing, for each point, the RSSI values corresponding to each AP present in the network; and another one to store the RSSI values corresponding to each AP taking also into account the points on which such AP appears[19].

The database is also populated with pre-calculated values, such as mode, average, maximum, minimum, mean value, and the analysis of the inner quartiles of the samples [19].

After building the database, vectors are obtained for each point, which are then processed to obtain the corresponding positions. The result of this process is a set of Euclidean distances between the sample vector and the database. Among these, the one with the lowest value is selected, which determines the position of the MD. This process is repeated for each sample vector, and is processed with each of the pre-processed values.

The tool also allows selecting the area to analyze, so that the APs that do not provide a good signal, due to their either not being near enough or their being too close, can be excluded. This option allows limiting the area from which information is obtained for processing. The RSSI values corresponding to APs that are either very far away or very close to de MD bias the measurement and reduce accuracy.

The offline mode is projected to allow the automatic analysis of the vectors corresponding to each sampled point. Processing output is stored in a flat file that can be analyzed at a later time. This can be used to decide which of these techniques are the ones that help the most to improve positioning accuracy.

Two structures were created for database storage – one that allows streamlining data pre-processing, and another one that speeds up calculations when positioning the MD.

The former, shown in Fig. 1, allows knowing all involved APs within the environment being analyzed.

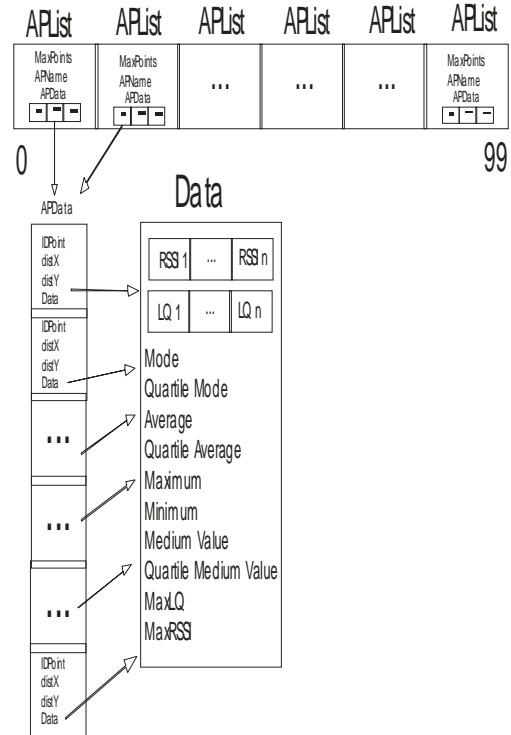

**Figure 1.** Structure by AP

The structure is an array of APs where name information is stored, and it includes a list of points that include the relevant AP. At each point, there is a structure that stores sampling data. There, the RSSI and LQ (Link Quality) data associated with that particular AP are stored. The data processed by the tool, such as mode, average, mean value, and so forth, are also stored. The purpose of this structure is making data processing simpler by going through the list of points for each AP that is present in the environment.

The second structure, depicted in Fig. 2, represents the environmental distribution used for the experiments carried out. A matrix is created, where each cell has the sampling data corresponding to that point. The RSSI and LQ data corresponding to each AP present at that point are stored there.

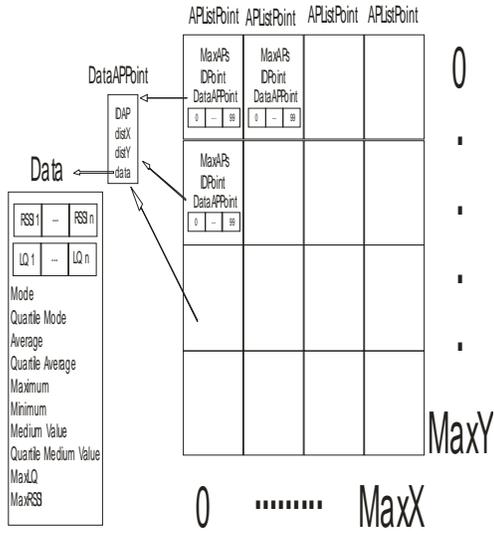

**Figure 2.** Structure by point

This type of storage allows for a simpler processing of the data and speeds up calculations for MD positioning in real time.

The structures are designed to be read as a double-entry table: by AP and by matrix point. Thus, a data mesh is formed that can be accessed based on the needs of the process to be carried out. First, when processing data and calculating the values corresponding to the mode, average, maximum, minimum, etc., the structure that is organized by AP is used, since it provides a simple and quick way for storing the data. Then, when the MD is to be positioned, the second structure is used, since it is organized following the points of the matrix that is built for positioning and that represents the environment from where the data were taken. This second structure speeds up the calculations performed to determine the position of the MD, thus requiring less time than what would be needed if using the technique without the tool.

## IV.     Experiment

Experiments were carried out at the library of the research institute INTIA/INCA, School of Exact Sciences, National University of the Center of the Province of Buenos Aires, Tandil, Argentina. The area spans over approximately 36 $m^2$. To carry out measurements and design the radio map, the area is divided into 100-cm positions along orthogonal coordinates (row, column), as shown in Fig. 3.

The Fingerprint database is built with 36 sampling points. At each point, approximately 2 minutes of samples are collected, which translates into more than 100 individual data per point. Therefore, a base with more than 3600 samples is obtained; these samples become the database used for calculating Euclidean distances. Mode, average, maximum, minimum, etc., values are added to streamline the process

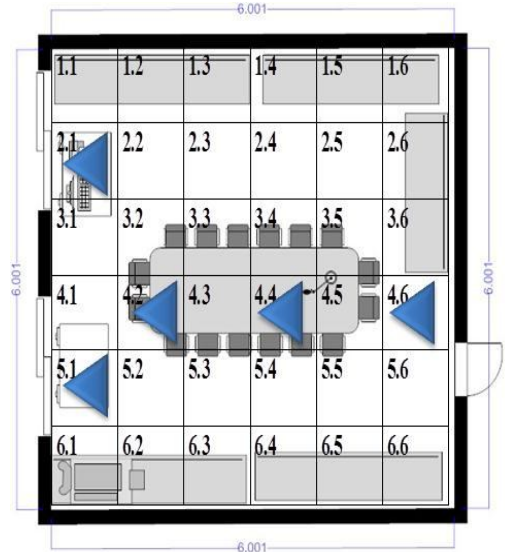

to position the MD when in the online mode.
**Figure 3.** Coordinates location in the map.

## V.     Result Analysis

During the calculation phase, the MD is positioned and "pattern vectors" are obtained for the strengths of the visible APs for a period of 120 seconds. Then, with these vectors and each of the strength vectors previously stored (maximum values, minimum values, average, mode, and interquartile pair mean value), the Euclidean distance is calculated by obtaining the distances to each coordinates point.

The position of the MD is determined as the pair of coordinates associated to the lowest value that meets the equation, that is, the shortest distance between the training set obtained (Fingerprint database) and the input data pattern.

Multiple tests are carried out and the analysis of errors is sorted as follows: first, a quantitative analysis of the hits obtained with each technique is performed.

Secondly, considering the actual position of the device and the position calculated by each of the techniques, it is possible to determine the maximum error to ensure an accuracy of 95% by implementing the techniques.

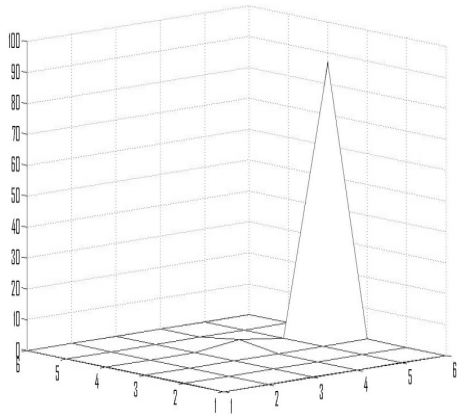

**Figure 4.** Surface chart for point 4.6.

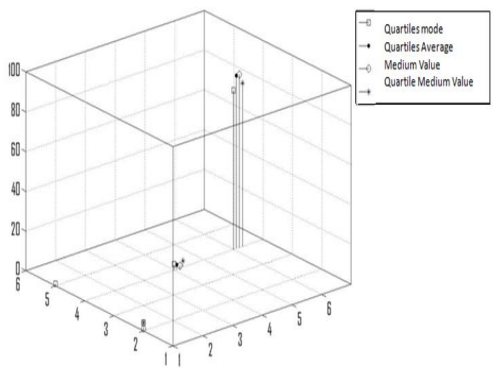

**Figure 5.** Distribution chart for point 4.6.

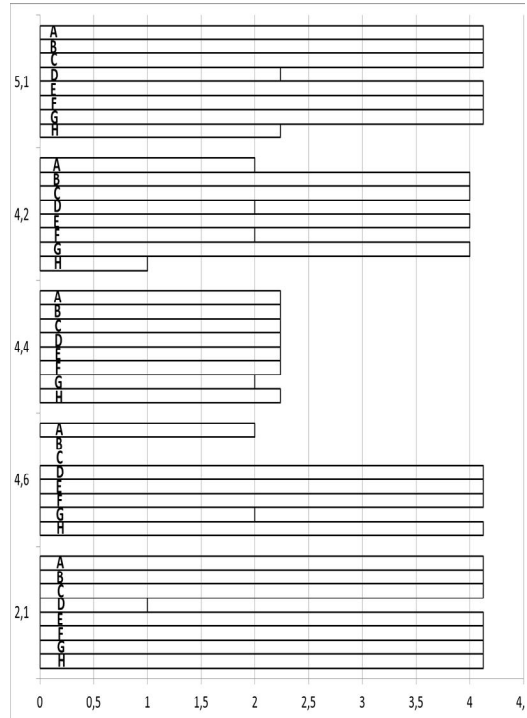

**Figure 6.** Error analysis, in meters. A: Maximum, B: Minimum, C: Mode, D: Quartiles Mode, E: Average, F: Quartiles Average, G: Mean Value, H: Quartiles Mean Value.

Fig. 4 presents a surface chart with the quartile average technique for position 4.6. As it can be observed, 87 vectors indicate that the MD is positioned at point 4.6, while 2 vectors yield point 4.4 as their result. There is a hit percentage of 98%.

Fig. 5 shows a distribution of the positions calculated for the vectors obtained at position 4.6. The percentages of hits for the different techniques are: quartile mode 89%, quartile average 98%, mean value 98%, and quartile mean value 92%. At points 4.4, 2.1 and 5.1 there are false positives which, in the worst of cases, correspond to 0.4% with the "quartile mode" and "mean value" techniques.

Fig. 6 presents the maximum errors (measured in meters) with each technique for the five positions used during the calculation phase. As it can be seen, the inner positions have a maximum error of 2.2 meters, whereas in the remaining positions, that error is 4.2 meters. This larger error is caused by signal strength fluctuations and signal absorption on obstacles that are adjacent to the sampling points.

At positions 4.2, 2.1, on the ends, the "average" technique achieves the lowest error margin, which in the worst of cases is 2.2 meters, reducing by half the errors obtained with the remaining techniques.

## VI. Performance Metrics

The WiFiPos tool allows calculating the position of the MD by calculating the Euclidean distance between the vector obtained when positioning the device and the pre-processed values of the different techniques.

If there are 100 vectors with RSSI values for the various APs, the time required to position a MD is reduced by a factor of 100. This is because the data collected are pre-processed offline during the previous stage, as opposed to the traditional Fingerprint technique that processes each datum as it is entered into the database Thus, the information obtained is simplified by pre-processing it offline.

With this proposal, the time required to position a MD can be significantly reduced when the system is working online. Reducing positioning time is required because the system is working under the constraints of real time device positioning.

To achieve an acceptable response time, the structures presented in the previous sections were created. These are designed to allow meeting the demands of response times. On the one hand, Structure 1 can store the data acquired from each AP and obtain values such as mode, average, maximum, minimum, and so forth. Thus, data are pre-processed and available when the positioning request arrives. This is possible because this structure is organized as a list of APs with their respective values. After processing the data, Structure 2 is simultaneously loaded.

On the other hand, Structure 2 reorganizes the data acquired and pre-processed to optimize calculations when the position of the MD is to be obtained, since it is organized based on the point from which samples were gathered. This way of storing the data allows reducing response time when the request to position the MD is received.

## VII. Conclusions and Future Works

The analysis of the data using the tool and applying various techniques allowed improving accuracy while reducing response times.

The results obtained when analyzing the data with the tool presented show that 95% of the times, the position of the MD can be determined with a relative error of 4.2 meters. It should be noted that this analysis was carried out over all errors, the rest being hits. Hits are those cases when the position calculated by the tool matches that of the data gathered. Anything else is considered as an error.

The addition of a set of additional techniques is planned, as well as adding the ability of automatically selecting the best technique to determine the position.

The tool is designed to process data from other sources or sensors, which will allow adjusting the position by analyzing and processing other signals. Among these, Bluetooth for indoor spaces and GPS for outdoor spaces are included.

## VIII. ACKNOWLEDGMENTS

This work was supported by Consejo Nacional de Investigaciones Científicas y Técnicas.